# Thin Film Lithium Niobate on Diamond (LiNDa) platform for Efficient Spin-Phonon Coupling

Zhujing Xu[1*], Sophie Weiyi Ding[1,2], Eliza Cornell[1], Salma Mohideen[1,3], Matthew Yeh[1], Kazuhiro Kuruma[1,4], Leticia Magalhaes[1], Amirhassan Shams-Ansari[1,5], Benjamin Pingault[6,7] and Marko Lončar[1*]

[1] *John A. Paulson School of Engineering and Applied Sciences, Harvard University, Cambridge, Massachusetts, USA.*
[2] *AWS Center for Quantum Computing, San Francisco, California, USA.*
[3] *Department of Physics, Stanford University, Stanford, California, USA.*
[4] *Research Center for Advanced Science and Technology, The University of Tokyo, Tokyo, Japan.*
[5] *DRS Daylight Solutions, 16465 Via Esprillo, San Diego, California, USA.*
[6] *Center for Molecular Engineering and Materials Science Division, Argonne National Laboratory, Lemont, Illinois, USA.*
[7] *Pritzker School of Molecular Engineering, University of Chicago, Chicago, Illinois, USA.*
*Corresponding author. Email address: zxu@seas.harvard.edu; loncar@g.harvard.edu

**Abstract**

Negatively charged silicon vacancy (SiV⁻) center in diamonds are leading candidates for solid-state quantum memories that can be controlled using electromagnetic or acoustic waves. The latter are particularly promising due to strong strain response of SiV⁻, enabling large spin-phonon interaction strengths. Indeed, coherent spin control via surface acoustic waves (SAW) has been demonstrated and is essential for developing on-chip phononic quantum networks. However, the absence of piezoelectricity in diamond requires interfacing with a piezoelectric material for efficient transduction and delivery of acoustic waves to spins in diamonds. Here, we demonstrate a heterogeneously integrated phononic platform that combines thin-film lithium niobate (TFLN) with diamond to enable acoustic control of single SiV⁻ spins. Additionally, leveraging large SAW-induced strain at the location of SiV⁻, we achieve coherent acoustic control of an electron spin with more than twofold improvement in Rabi frequency compared to the state-of-the-art devices based on aluminum nitride-on-diamond. This work represents a crucial step towards realizing phonon-based quantum information processing systems.

**Introduction**

Recent progress in quantum acoustics has advanced the development of phonon-mediated quantum systems across a range of platforms, including superconducting qubits(*1–8*), solid-state quantum emitters(*9–13*), quantum dots(*14–16*), and photonics(*17–19*). Among these platforms, negatively charged silicon vacancy (SiV⁻) centers in diamonds have emerged as promising candidates for long-lived quantum memories and scalable quantum network nodes(*20, 21*). In addition, SiV⁻ spins exhibit high strain susceptibility(*22, 23*), enabling efficient coupling to phonons(*24, 25*). Notably, acoustic waves have been shown to coherently drive both electron spins and nuclear spins in negatively charged silicon vacancy (SiV⁻) centers in diamonds, achieving coupling efficiencies that are two orders of magnitude higher than those obtained via direct microwave driving(*10, 26*). These demonstrations highlight the unique advantage of mechanical degrees of freedom in quantum information processing.

However, such demonstrations rely on interfacing diamond with a piezoelectric material for efficient excitation of SAWs, as diamond itself lacks piezoelectric properties. Among the available materials, single-crystal TFLN stands out due to its strong piezoelectricity ($d_{24} = d_{15} \sim 70\ pC/N$(*27*)) and low acoustic loss which have been leveraged to demonstrate high-frequency mechanical resonators(*28–30*), high quality factor MEMs resonators(*31, 32*), acoustic delay line oscillators(*33–35*), quantum transducers(*18, 36*), integrated phononic circuits(*37, 38*), acousto-optic modulator(*39, 40*) and beam-steering systems(*41–44*). Furthermore, TFLN possesses multiple non-zero piezoelectric tensor components, enabling the generation of strong and controllable strain fields that can enhance spin-phonon coupling in hybrid quantum systems. Although TFLN demonstrates significant promise for quantum acoustics, its integration with diamond has so far been limited to photonic applications(*45*), while its potential for high-performance acoustic devices remains unexplored.

In this work, we demonstrate the heterogeneous integration of thin film lithium niobate on diamond (LiNDa), a new platform for the realization of high-performance SAW quantum devices (Fig.1(C)). Our approach relies on a micro-transfer printing(*46*) process to place a TFLN membrane onto the diamond substrate, followed by standard microfabrication processes to fabricate interdigitated transducers (IDTs) on the LN device layer. Utilizing this heterogeneously integrated LiNDa platform, we realize a bidirectional SAW device exhibiting a central resonance frequency of approximately 3.8 GHz and an intrinsic quality factor of 2,450. Leveraging this, we demonstrate a more than twofold improvement of Rabi frequency compared to previously reported results(*10*), currently limited by unoptimized spin placement and contamination in the cryostat. Our results demonstrate a crucial step toward the incorporation of spins in diamond and integrated phononic platforms. The LiNDa platform provides efficient bidirectional conversion between microwave and acoustic signals, which shows its promise as a phononic link connecting a superconducting quantum processor to spin-based quantum memories in solid-state materials(*47*). Besides, this novel structure holds substantial potential for meeting the increasing demands for high-frequency, large-bandwidth RF front-end filters, particularly relevant to the emerging 5G communication standards(*48, 49*).

**Results**

Our SAW devices are fabricated on electronic grade single-crystal diamond (Element Six) with a 600-nm layer of x-cut TFLN bonded on top. We chose the z-axis of LN as the direction of propagation of SAW. We start our fabrication by releasing TFLN membrane, with a size of 250 $\mu m$ by 200 $\mu m$, from commercial TFLN wafer (NanoLN). The TFLN membrane is then transferred to a diamond substrate by a micro transfer printing technique(*46*). The TFLN and diamond are strongly bonded by the Van der Waals force. Finally, the metal IDTs, used to generate and detect acoustic waves, are defined on the TFLN using electron beam lithography (EBL) followed by metal deposition and liftoff (fabrication details in supplementary materials). The spacings between IDT fingers, which sets the acoustic wavelength, is chosen to be 1.7 $\mu m$ corresponding to a generated SAW frequency of 3.8 GHz. The transducers efficiently convert electrical signals to acoustic signals and can excite all five fundamental acoustic modes in this heterogeneous structure, including two shearing modes and three Rayleigh modes. The profiles of supported SAW modes, obtained using finite element modeling, are shown in Fig. 1B.

The devices are first characterized with a vector network analyzer. Our frequency-domain measurement of transmission and reflection for a device with a propagation length of 30 $\mu m$ (as shown in Fig. 1C) reveals all five modes. The third resonant peak, corresponding to one of the Rayleigh modes, shows the highest transmission due to good impedance matching and the highest electromechanical coupling efficiency among all fundamental modes (Fig. 1C). The phase velocity of the Rayleigh mode is measured to be about 6477 m/s, in good agreement with simulated value 6990 m/s. The slight discrepancy is due to fabrication uncertainties in the waveguide orientation, IDT spacings, and TFLN thickness. The 3dB bandwidth of the Rayleigh mode is greater than 300 MHz and is determined by the number of fingers used in the IDT. The maximum transmission $S_{21}$ of the Rayleigh mode is measured to be -10.7 dB at 3.8 GHz, over a 30 $\mu m$ propagation length, which is 10 dB better than the SAW transmission previously reported in aluminum nitride on diamond(*10*). The high transmission comes from the low acoustic loss and high electromechanical efficiency of TFLN, as well as low-loss interface between LN and diamond, confirming the quality of our heterogeneous integration. We note that, unlike photonics examples of micro-transfer printing, our method does not use any intermediate layer such as benzocyclobutene (BCB)(*50–52*) to enhance adhesion between the LN and diamond, relying solely on a clean and properly terminated interface and Van der Waals forces. The transmission through the IDTs is theoretically limited to -6 dB, as the generated acoustic waves are launched symmetrically in both directions. This constraint can be overcome by employing unidirectional transducers (UDTs) that preferentially launch waves in a single direction(*53–55*).

Since the aluminum IDTs can serve as reflective mirrors for propagating SAWs, an acoustic cavity is formed between the two IDTs. The reflection spectrum in Fig. 2A reveals a series of evenly spaced cavity modes with a free spectral range (FSR) of 52.6 MHz, for a device with a longer acoustic cavity with d = 50 μm. This allows us to evaluate the effective length of the cavity and determine the penetration depth of the SAW into each IDT. We find a penetration the depth to be $L_p$ = 4.3 μm, which corresponds to measured reflectivity of $|r_s| \sim 10\%$ for each electrode in the IDT(*56*). A SAW cavity confined by metal mirrors typically has weak confinement due to the small velocity contrast between metal-clad and unclad regions(*57*). Consequently, achieving high reflectivity often a large number of IDTs, resulting in increased device footprint and enhanced Ohmic losses. In contrast, the metal mirrors implemented on the LiNDa platform provide significantly stronger confinement with the reflectivity of each electrode exceeding that of the metal Fabry-Perot cavity on quartz(*56*), the cavity with etched grooves in GaN-on-sapphire(*58*) and the cavity with etched grooves in GaAs(*15*). The enhanced reflectivity comes from the strong electromechanical coupling of TFLN and the high acoustic index contrast between TFLN and diamond. This enables a reduction in the number of metal mirrors, thereby minimizing ohmic losses and simplifying the fabrication process.

We fit the reflection spectrum with a Lorentzian function and find the overall quality factor be $Q \sim 2100$, which corresponds to an intrinsic quality factor of $Q_i \sim 2450$ (Fig. 2B). The intrinsic quality factor ($Q_i$) for all modes shows an increasing trend versus their resonant frequencies (Fig. 2C). This trend in $Q_i$ can be attributed to reduced bulk scattering losses as the cavity mode reaches the high-frequency edge of the mirror's stop band where the group velocity is slower(*15*). With the reflectivity of each electrode, we can calculate the scattering loss through the mirror and find the quality factor limited by mirror scattering is $Q_{mirror} = \frac{\pi(d+L_p)}{\lambda_0(1-\tanh(N_{mirror}|r_s|))} = 1.61 \times 10^5$. Here $d = 50$ μm is the distance between two IDTs, and $N_{mirror} = 40$ is the number of mirrors in the Fabry-Perot cavity. The acoustic propagation loss of the LiNDa platform is extracted to be $\alpha_p = 3.2$ dB/mm using a time-domain measurement (details on propagation loss calibration in supplementary materials). Therefore, the quality factor limited by propagation loss of our Fabry-Perot cavity is $Q_{propagation} = \frac{\omega}{2v_g\alpha_p} = 2638$, where $v_g = 6161$ m/s is the group velocity. This gives the cavity's intrinsic quality factor $Q_i = (Q_{propagation}^{-1} + Q_{mirror}^{-1})^{-1} = 2595(56)$, which is similar to the measured $Q_i$ in the experiments. With the current design, the quality factor is not limited by the electrode number, but instead is limited by materials loss, diffraction of spreading acoustic wave and radiation loss to the bottom substrate. The finesse of the cavity at 5K is $F = Q_{total}\lambda/(2(d + 2L_p)) = 30.5$, which is comparable to the SAW cavity in GaAs at mK temperature(*15*). Improved device performance can be achieved through better device design and material treatments.

To reduce the SAW transverse profile and thus enhance the spin-phonon interaction, we use curved IDTs that can generate a Gaussian SAW beam which focuses the acoustic energy to a waist (Fig. 3A). The SAW mode is evanescently coupled to a SiV⁻ spin implanted around 50 nm beneath the diamond surface via the strain it generates. We design a three-port device consisting of two IDTs along the TFLN z-axis and a single IDT along the TFLN y-axis. The three-port device is chosen to allow for future experiments that require two simultaneous SAW drives, a continuous-wave dressing of the SiV⁻ spin and a pulsed probing of the dressed spin state(*59*). In this work we mainly focus on the single-sided device to generate coherent control pulses needed for the efficient coherent control of a single SiV⁻ spin. The number of electrodes in each IDTs is chosen to match the 50 ohm impedance of the source electronics. The microwave measurements of the Gaussian SAW devices at room temperature are shown in Fig. 3B and Fig. 3C. The center frequency of the z-axis device is around 4.1 GHz, and the center frequency of the y-axis device is around 3.8 GHz. The wavelengths of the z-axis device and y-axis device are 1.7 $\mu m$ and 1.1 $\mu m$, respectively. The wavelengths are chosen such that the SiV⁻ ground-state spin splitting (Fig. 3D) can be Zeeman-tuned into resonance with the mode frequencies via nearby permanent magnet. By adjusting the magnet position, the spin splitting of a single SiV⁻ can be matched to the acoustic mode frequency and the spin can be resonantly driven, as was first demonstrated in (*10*).

Acoustic measurements of a single SiV⁻ spin are performed in a closed-cycle liquid helium cryostat whose cold stage temperature is measured to be 4.7 K. We use a tunable 737 nm laser to resonantly excite the optical transitions of individual SiV⁻ centers. When a continuous acoustic wave reaches the SiV⁻ spin, the modulation of the optical energy results in sidebands in the resonant excitation spectrum. Fig. 3E shows the sidebands of the C transition arising from a continuous acoustic wave of 3.83 GHz with 0 dBm power at the input to the cryostat. Each pari of the measured peaks are well fitted by a double Lorentzian function, corresponding to the C2 and C3 transitions.

We find the SiV⁻ spin resonance frequency and confirm its interaction with the SAW mode by taking an optically detected acoustic resonance (ODAR) measurement, whose pulse sequency is illustrated in the inset in Fig. 4A. The optical pulses are generated by an acousto-optic modulator and the microwave pulses driving the IDTs come from an arbitrary waveform generator (AWG). The laser is tuned resonant with C3, such that the fluorescence counts in the first part of the optical pulse are proportional to the population in the $|\uparrow\rangle$ state. The first 300 ns optical pulse also serves to initialize the spin by optically pumping the population into the $|\downarrow\rangle$ state. It is followed by a 20-ns microwave SAW pulse which drives the spin. Finally, another 300 ns optical pulse is applied to read out the final population, and the relative population is determined by the ratio between the signals from the two pulses. The SAW frequency is swept, and when the strain at the SiV⁻ is modulated at a frequency matching the spin splitting between the $|\downarrow\rangle$ and $|\uparrow\rangle$ states, it resonantly drives the population between two states. The resulting ODAR peak is shown in Fig. 4B, indicating that the resonance frequency of the SiV⁻ ground state spin splitting is 3.83 GHz.

We then fix the SAW frequency at 3.83 GHz and vary the SAW pulse duration length to measure the Rabi oscillations (Fig. 4B). As the duration length of the SAW pulse increases, we observe a coherent oscillation of population between the $|\downarrow\rangle$ and $|\uparrow\rangle$ states. The Rabi measurement is well fitted with an exponentially decaying sinusoidal function, giving a Rabi frequency of 33.4 MHz for an input microwave power of -4 dBm. As expected, we observe that the Rabi frequency increases linearly with the square root of the power increases, as shown in Fig. 4C. The LiNDa platform achieves of more than twice of Rabi frequency of the current state-of-the-art AlN-on-diamond platform, when compared at the same applied microwave power(*10*).

**Discussion**

We attribute this more efficient Rabi driving to TFLN's higher microwave-to-acosutic transduction efficiency and the larger single-phonon strain enabled by the SAW mode geometry. Our modeling reveals that the electromechanical coupling efficiency ($k^2$) of the SAW mode generated by the Gaussian IDTs is 25% (Fig. 3). Our model on the strain generated by a single phonon indicates that for a SiV⁻ spin located 50 nm beneath the diamond surface, the single spin-phonon coupling rate is $g = 30\text{-}70$ kHz for different SiV⁻ orientations. Here the waist of the Gaussian IDTs is $w_0 = 6.8\ \mu m$. More details about the simulations can be found in the supplementary materials.

For comparison with previous SAW studies, the key figure of merit is the maximum achievable Rabi frequency for a given microwave driving power. The transmission measurement indicates -10 dB loss from RF-to-acoustic conversion and material loss. In addition, -10 dB insertion loss arises from wire bonding and other electrical connections. The power of a single phonon is given by $p_0 = \hbar\omega_0/t_0 = 1.25 \times 10^{-16} W$, where $t_0 = 20\ ns$ is the duration of the phonon calculated from the measured bandwidth of the SAW device (around 50 MHz)(*60*) and $\omega_0 = 2\pi \times 3.8 \times 10^9$ Hz is the frequency of the phonon. In this case, 1 mW of RF power results in 10 $\mu$W acoustic power, which corresponds to $n = 7.98 \times 10^{10}$ phonons in the system. Therefore, the corresponding total Rabi rate would be $\sqrt{n}g = 8.5\text{-}20$ GHz. This is more than 150 times larger than the Rabi rate we measured with the same power in our experiments. This large discrepancy is attributed to unoptimized placement of SiV⁻ (not at the center of the Gaussian waist) and significant oil contamination in our cryostat resulting in degraded SAW devices (details in supplementary materials). Further improvement can be done with better design like UDTs(*53–55*), more focused Gaussian IDTs, more optimized SiV⁻ location, and better cryostat systems.

The LiNDa platform significantly outperforms the AlN-on-diamond platform which has in recent years been the workhorse of diamond SAW experiments(*10, 26, 60*). Among the SAW modes supported by AlN-on-diamond, the Rayleigh mode exhibits the highest electromechanical coupling coefficient of $k^2 = 1.3\%$ (*60, 61*). At the same operating frequency of 3.8 GHz, the corresponding single spin-phonon coupling rate for a Rayleigh mode on AlN-on-diamond structure is 8-13 kHz with the same mode volume as mentioned above. As a result, the LiNDa platform offers a 19-fold enhancement in $k^2$ and 5-fold increase in strain response, which together could enable over a 20-fold increase in the achievable Rabi frequency.

In conclusion, we present the heterogeneous acoustic integration of lithium niobate and diamond for enhancing the spin-phonon coupling of SiV⁻ color centers. We have demonstrated this platform's improvements in its low acoustic loss and the efficient acoustic coherent control of a single SiV⁻ spin. In theory, the single spin-phonon coupling strength can be enhanced by more than 20 times of magnitude, compared to previously reported results(*10*). Moreover, the strong electromechanical coupling of TFLN, combined with the high acoustic index contrast between TFLN and diamond, enables metal electrodes on the platform to work as reflective acoustic mirrors. This yields a cavity with an intrinsic quality factor of 2,450, which is important for enhancing the spin-phonon interaction even further. The LiNDa platform offers a promising route towards quantum acoustodynamics (QAD) and phonon-mediated interfaces across various quantum systems.

**Acknowledgments**

**Funding:** We would like to thank Andrea Cordaro, CJ Xin, Alex Raun and Danial Haei for their helpful discussion. Z.X. acknowledges support from Harvard Quantum initiative (HQI) postdoctoral fellowship. M.Y. acknowledges support from the Department of Defense (DoD) through the National Defense Science and Engineering Graduate (NDSEG) Fellowship Program. L.M. acknowledges funding from Behring foundation and CAPES-Fulbright. Z.X., S.W., and E.C. acknowledges support from NSF Engineering Research Center for Quantum Networks (EEC-1941583), and Air Force Office of Scientific Research (AFOSR) MURI on Quantum Phononics. Device fabrication was performed at the Harvard University Center for Nanoscale Systems (CNS), a member of the National Nanotechnology Coordinated Infrastructure Network (NNCI), which is supported by the National Science Foundation under NSF award no. 1541959.

**Author contributions:** Z.X, S.W.D, A.S, K.K, and M.L conceived the idea for the project, Z.X, S.W.D, S.M, A.S, and K.K fabricated devices, Z.X, E.C and M.Y performed the experiments, Z.X and L.M performed the simulations, Z.X, E.C, M.Y and B.P analyzed the data, M.L. supervised the project. All authors contributed to the writing of the manuscript.

**Competing interests:** Authors declare that they have no competing interests.

**Data and materials availability:** All data needed to evaluate the conclusions in the paper are present in the paper and/or the Supplementary Materials.

# Figures and Tables

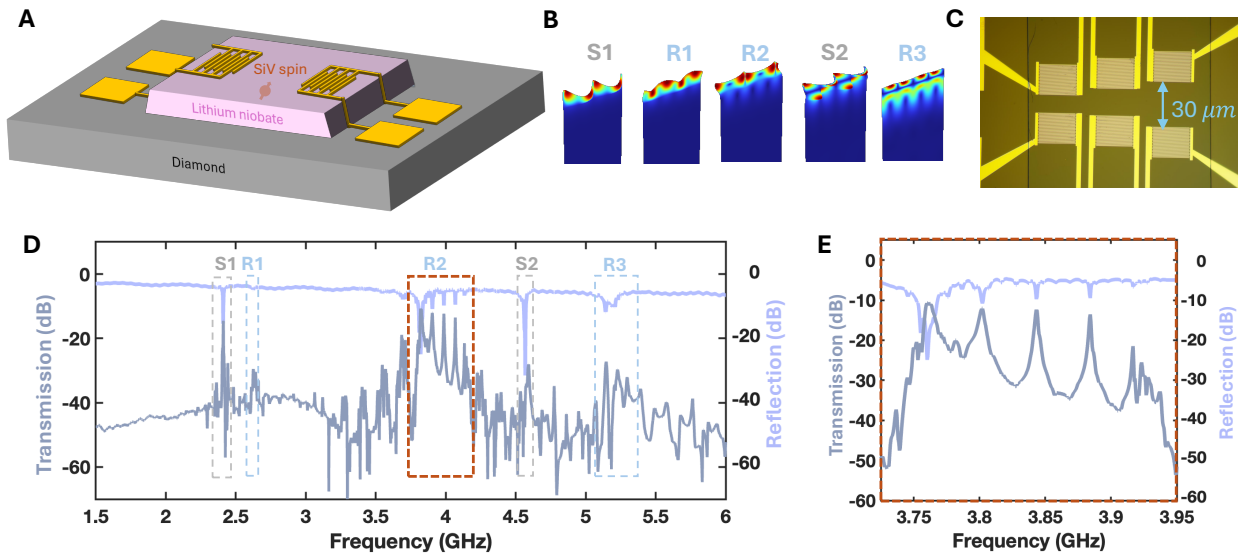

**Fig. 1. Surface acoustic wave (SAW) devices on a heterogeneously integrated thin film lithium niobate on diamond platform.** (**A**) Schematic of a SAW device realized in a heterogeneously integrated thin film lithium niobate and diamond (LiNDa) platform. A 600-nm thick x-cut TFLN layer is transferred onto the diamond and bonded by van der Waals force. Aluminum interdigital transducers (IDTs) are patterned on top of the TFLN to efficiently transduce acoustic waves. (**B**) Cross-sections of the simulated SAW modes. There are five fundamental modes, including two shearing modes (S1 and S2) and three Rayleigh modes (R1, R2 and R3). (**C**) An optical image of the SAW device on this heterogeneous platform. The SAW propagates along the z-direction of the LN. The IDTs launch the acoustic wave bidirectionally and serve as reflective mirrors for acoustic waves. (**D**) Frequency-domain measurement of a SAW device with a 30-$\mu$m long device (length is defined as the separation between the two IDTs) as shown in (**B**). The device supports all five fundamental SAW modes whose resonant frequencies are 2.41 GHz, 2.62 GHz, 3.82 GHz, 4.57 GHz, and 5.14 GHz, while the theory predicts the five fundamental modes at 2.69 GHz, 2.93 GHz, 4.17 GHz, 5.15 GHz, and 5.94 GHz. The frequency difference comes from fabrication imperfections, thin-film lithium niobate thickness variations and mass loading effects from the IDTs. (**E**) A zoom-in showing the R2 mode. The maximum transmission $S_{21}$ is -10.7 dB at 3.8 GHz and the 3dB bandwidth is larger than 300 MHz. The reflection and transmission measurement ($S_{11}$ and $S_{21}$) reveal a series of evenly spaced cavity modes which are formed by the reflections between two sets of IDTs.

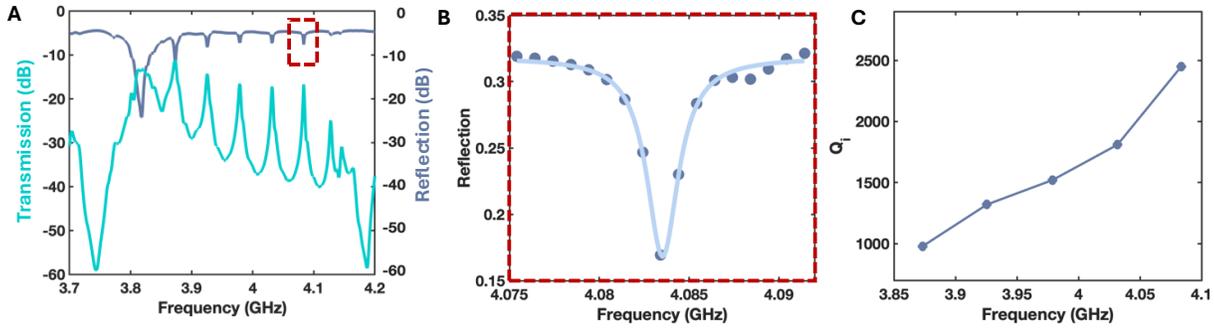

**Fig. 2. Microwave characterization of planar SAW cavities.** (**A**) Frequency-domain measurement of a 50-um long SAW device. The aluminum IDT electrodes serve as reflective mirrors forming a Fabry-Perot cavity. A series of evenly spaced cavity modes are observed, and the free spectral range is measured to be FSR = 52.6 MHz. (**B**) The cavity mode at 4.08 GHz is fitted with a Lorentzian function resulting in an overall Q factor of 2100. This corresponds to an intrinsic $Q_i$ of 2450. (**C**) The intrinsic $Q_i$ of each cavity mode vs the mode's resonant frequency. The increase in $Q_i$ is attributed to reduced bulk scattering losses as the cavity mode reaches the high-frequency edge of the mirror's stop band.

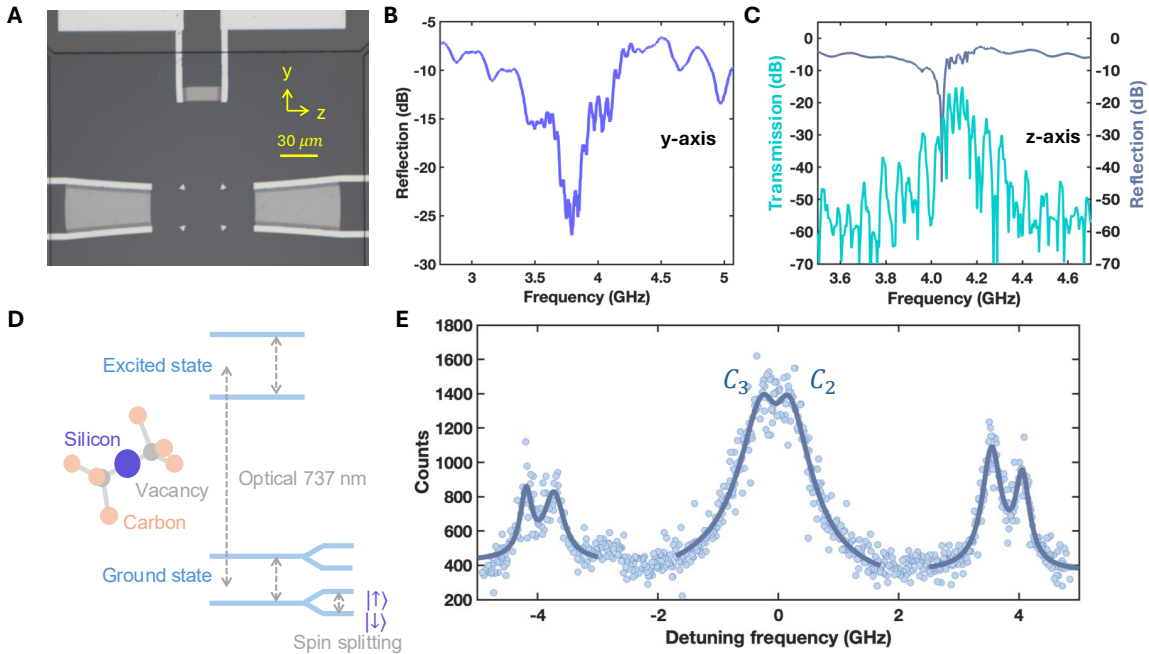

**Fig. 3. Microwave characterization of Gaussian SAW devices for coherent control of SiV⁻ spins. (A)** Schematic of three-port device used for coherent control of SiV⁻ electron spins. We use two IDTs along the z axis of LN and a single IDT along the y axis. **(B)** Reflection measurement of the single IDT device along the y-axis of LN. The center frequency is around 3.8 GHz. **(C)** Transmission and reflection measurement of the IDT devices along the z-axis of LN. The center frequency is around 4.1 GHz. **(D)** Electronic energy level of a SiV⁻ spin in diamond. The spin splitting in the ground state is tunable by an external magnetic field. The optical transition between the ground state and the excited state can be used for initialization and readout for the spin states $|\downarrow\rangle$ and $|\uparrow\rangle$. Acoustic waves can be used to drive the transitions between two spin states. **(E)** SAW modulation-induced sidebands of C transition of an SiV⁻ spin's C transition when 0 dBm of RF power at 3.83 GHz is applied to the y-axis IDT. The measurement is well-fitted by a double Lorentzian function.

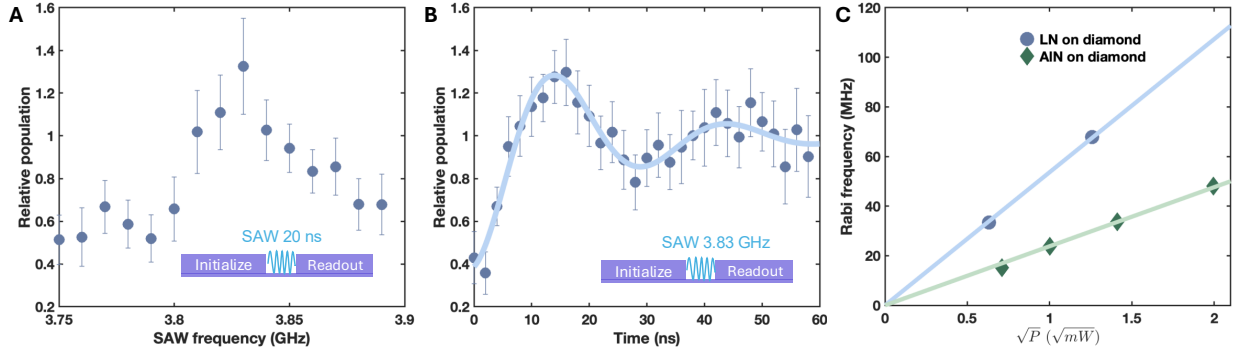

**Fig. 4. Coherent acoustic control of a single SiV spin. (A)** Optically detectable acoustic resonance (ODAR) measurements. The spin is first initialized to the $|\downarrow\rangle$ state by an optical initialization pulse. This is followed by a 20-ns SAW pulse which flips the state when it si resonant with the spin transition frequency. After this, the population of the $|\uparrow\rangle$ state is read out by another optical pulse. The normalized population in the $|\uparrow\rangle$ state is shown versus frequency of the acoustic pulse. The ODAR peak at 3.83 GHz shows the resonance frequency of the SiV⁻ spin transition. **(B)** Rabi measurements. The spin is first initialized to the $|\downarrow\rangle$ state. The normalized population in the $|\uparrow\rangle$ state is measured as the acoustic pulse duration length is varied while its frequency is fixed at 3.83 GHz. The measured Rabi frequency is 33.4 MHz with a microwave power of -4 dBm at the input of the cryostat. **(C)** The power dependence of Rabi frequency reveals a linear behavior. Our measurements show that the LiNDa platform shows a more than twofold improvement in Rabi rate, compared to the previously reported results for the aluminum nitride on diamond platform(*10*).

# Supplementary Materials for

# Thin Film Lithium Niobate on Diamond (LiNDa) platform for Efficient Spin-Phonon Coupling

Zhujing Xu *et al.*
Corresponding author. Zhujing Xu, *zxu@seas.harvard.edu;* Marko Loncar*, loncar@g.harvard.edu*

1. **Device fabrication**

   Thin-film lithium niobate (TFLN) has been patterned on various substrates, including silicon(*62*), silicon carbide, silicon dioxide(*63*) and sapphire(*64*). However, TFLN on diamond (LiNDa) samples have not been reported yet. In this work, we pattern and undercut a TFLN membrane on a LN-on-insulator sample and transfer print it onto a bulk diamond chip. The transfer printing process used in this work is illustrated in Fig. S1(A). The fabrication starts with patterning a 600-nm thick X-cut LN on a 4.7-$\mu m$ thick thermal silica buffer layers on a 525-$\mu m$ thick silicon handle wafer. The patterns are defined by photolithography with SPR700-1.0 as the resist. The patterns are transferred to the LN with argon ion etching. The photoresist is then stripped in a piranha solution (3:1 98% sulfuric acid and 40% hydrogen peroxide). The LN membrane is released by etching the silica in hydrofluoric acid (49% HF). The suspended LN membranes are finally cleaned in acetone and methanol and dried in a critical point dryer. An optical image of a suspended LN membrane is shown in Fig. S1(B).

   To prepare the polydimethylsiloxane (PDMS) stamps, we start with patterning the chrome photomask. Then we spin 50-$\mu m$ thick SU8 photoresist onto a silicon wafer. The pattern is transferred to the SU8 layer using contact photolithography. The base and the curing agent are mixed, poured into the SU8 mask, and baked at 120 degrees C for 20 mins. The patterned PDMS stamps on a silicon wafer are shown in Fig. S1(C). Each stamp is about 50- $\mu m$ thick and its lateral dimensions are 295 $\mu m$ by 463 $\mu m$ so that it is large enough to cover the whole suspended LN membrane area and break the tethers when lifting up the PDMS (Fig. S1(D)).

   To transfer LN membranes onto a diamond substrate, we cut a piece of patterned PDMS from the silicon wafer and attach it to a glass-slide. The glass slide is mounted on a set of translational stages and is lowered until the PDMS stamp fully covers the LN membrane. Due to the viscoelasticity of PDMS, rapidly lifting the PDMS away from the sample causes the suspended LN membranes to detach from their tethers and adhere to the PDMS. We then bring the PDMS stamp with an attached LN membrane into contact with a single-crystal diamond chip. While the LN is touching the diamond, we heat up the diamond chip to 100 degrees Celsius for 10 mins. By slowly lifting the PDMS stamp from the diamond substrate, the LN membranes detach from the PDMS and remain adhered to the diamond. This process enables the simultaneous transfer printing of more than ten LN membranes. Subsequently, the chip is cleaned in acetone and IPA. An optical image of an LN membrane transferred onto a diamond substrate is shown in Fig. S1(E). The dimensions of the membrane are 200 $\mu m$ by 250 $\mu m$.

   We then fabricate SAW devices on the LiNDa platform, as shown in Fig. 1(B) in the main text. The metal interdigital transducers (IDTs) are defined with aligned electron beam lithography (EBL) using PMMA 950 C4 as the resist. We evaporate 110 nm aluminum and create the electrodes via a lift-off process in remover PG at 80 degrees Celsius. The wirebond pads are defined with another EBL write and we evaporate 10 nm titanium and 300 nm gold before performing another lift-off process.

2. **Propagation loss calibration**

   To extract the conversion efficiency and propagation losses for this heterogeneous platform, we can take the inverse Fourier transform of $S_{21}(\omega)$ which gives the time-domain response $|h(\tau)|$ (*35, 37*). An example of a time-domain response is shown in Fig. S2(A), for a 130-$\mu$m long device measured at 5K and room temperature. We observe a series of peaks (highlighted as dots in Fig. S2(A)) separated by a round-trip time $\Delta\tau = 2L/v_g$, where $L$ is the propagation length between two sets of IDTs and $v_g$ is the group velocity of the acoustic wave. The first peak (not highlighted) corresponds the the direct electrical crosstalk between two IDTs. The first highlighted peak represents the conversion and injection of the acoustic wave by the IDTs, and the following peaks result from reflections off the electrodes. The $n^{th}$ peak $h_{max}(n)$ depends on the IDTs conversion efficiency, propagation loss and reflection coefficient. It can be written as $|h_{max}(n)|^2 = T^2 R^{2n} e^{-\alpha(2n+1)L}$, where $\alpha$ is the propagation loss, $T$ is the conversion efficiency of the IDT and $R$ is the mechanical power reflection coefficient(35, 37). Therefore, we can write $2\ln|h_{max}(n)| = 2 \ln \ln T + 2n \ln \ln R - \alpha(2n + 1)L$ (37). By applying a linear fit to the $\ln|h_{max}(n)|$ (Fig. S2(B) and Fig. S2(C)), we can get the propagation loss $\alpha = 35.2 \; dB/mm$ at room temperature and $\alpha = 3.2 \; dB/mm$ at 5K. The propagation loss is a few times higher than the equivalent losses for SAW devices on TFLN-on-sapphire and suspended TFLN platforms(35, 37). The higher loss can be attributed to the divergence of the acoustic waves since the acoustic waves are not guided. A high propagation loss was also observed in other studies when the acoustic wave is not confined in a waveguide(37). Besides, the TFLN here is not annealed which could also contribute to the acoustic loss.

### 3. Calculating single spin-phonon coupling and Rabi rate

We first briefly discuss the electromechanical coupling coefficient $k^2$ which describes the efficiency of converting electrical energy into mechanical energy and depends on the material properties, and it can be calculated from the phase velocity of the SAW mode when the electrode on the piezoelectric layer surface is electrically open or shorted using $k^2 = 2(v_{open} - v_{short})/v_{open}$. We use COMSOL to perform a finite element analysis to simulate the phase velocity for SAW modes propagating along 600-nm x cut LN on diamond, and the results are shown in Fig. S3. We assume SAW propagation along the y-axis of LN and focus on the first shear mode because it has the highest $k^2$. The eigenfrequency and the $k^2$ of the LiNDa material stack are modeled for different wavelengths. The y-axis device in Fig. 3(A) in the main text has a wavelength of 1.1 $\mu$m. The resonant frequency of this shear mode is measured to be 3.83 GHz, while the simulation predicts it to be 3.89 GHz (Fig. S3(B)). Our modeling reveals that the $k^2$ of the shear mode at this wavelength is 25% (Fig. S3(C)).

Our modeling indicates that for an SiV⁻ located 50 nm beneath the diamond surface, the shear mode corresponds to a single spin-phonon coupling rate of $g = \frac{2\gamma_s B_x}{\lambda_{SO}} \sqrt{(d_s(\epsilon_{xx} - \epsilon_{yy}) + f_s \epsilon_{zx})^2 + (-2d_s \epsilon_{xy} + f_s \epsilon_{yz})^2}$ = 30-70 kHz for different SiV⁻ orientations(22). Here $\gamma_s$= 14 GHz/T is the gyromagnetic ratio of the spin and $B_x$ is the transverse magnetic field along the x-axis of the SiV⁻. In the experiment, we put a permanent magnet underneath the sample and the applied magnetic field is roughly along the [001] direction of the diamond chip. The ground-state splitting depends on the static strain and magnetic field. In the low strain regime, the resonance condition $2\gamma_s B_z = \omega_m$ provides an approximate relation between the axial magnetic field component $B_z$ and the mechanical mode frequency $\omega_m = 2\pi \times 3.83$ GHz. Thereby, we can get the transverse magnetic field component $B_x$ roughly to be $2\gamma_s B_x = \omega_m \times \tan\theta$, where $\theta \sim 54.7$ degrees is the angle between the B field direction [001] and the SiV⁻ axis. The tensor component in the SiV⁻ basis due to a single phonon in the mechanical mode is $\epsilon_{ij}$, and the strain susceptibilities are $d_s = 1.3$ PHz/strain and $f_s = -1.7$ PHz/strain. The orbital splitting in the ground state is $\lambda_{SO} = 46$ GHz. Here the waist of the Gaussian IDTs is $w_0 = 6.8$ $\mu$m. Here we only consider $E_{gx}$ and $E_{gy}$ strain when calculating the spin-phonon coupling strength since these two strain components modify the orbital splitting and hence couple the two ground spin states. In contrast, $A_{1g}$ strain shifts the whole ground state manifold in a uniform way and does not couple the ground spin states(*10, 22*).

The calculated single spin-phonon rate gives a Rabi rate of 8.5-20 GHz at 1 mW RF power (mentioned in the main text). This is 150 times larger than the Rabi rate we measured with the same power in our experiments. This large discrepancy is attributed to unoptimized placement of SiV$^-$ (not at the center of the Gaussian waist) and significant oil contamination in our cryostat resulting in degraded SAW devices. Next, we discuss the impact of each factor separately.

For the device presented in Fig. 3 and Fig. 4, the strain distribution is described by the Gaussian function that $u(r,z) = u_{max}(\frac{w_0}{w(z)})exp(\frac{-r^2}{w(z)^2})$, where $u_{max}$ is the strain at the center of the Gaussian beam, $w_0 = 6.8\ \mu m$ is the waist radius, $r$ is the radial distance from the center axis of the beam, $z$ is the axial distance from the beam's focus, $w(z) = \omega_0\sqrt{1+(\frac{z}{z_R})^2}$ is the spot size and $z_R = \pi w_0^2/\lambda$ is the Rayleigh range. For the SiV$^-$ we measured in the experiment, $z$ is about 70 $\mu m$ and r is about 10 $\mu m$, and hence the strain is about 0.16 $u_{max}$. Besides, there was unexpected oil contamination in our cryostat, and this has induced significant loss in our SAW device. We estimate an additional -20 dB loss added to RF-to-acoustic conversion. These two factors could make the Rabi rate 100 times smaller.

### 4. Acoustic ridge waveguide on LiNDa

The heterogeneous integration of TFLN and diamond combines the advantages of **(1)** LN's strong piezoelectricity which enables bidirectional efficient microwave-to-acoustic conversion, and **(2)** the large mismatch between the acoustic velocity of LN (4-7 km/s) and the acoustic velocity of diamond (12-18 km/s), which gives strong confinement of acoustic waves. Compared to other phononic platforms proposed previously like GaN-on-sapphire(*58*), LN-on-sapphire(*37*), and AlN-on-diamond(*10*), our LN-on-diamond platform has the highest acoustic refractive index contrast.

We designed and tested an acoustic ridge waveguide on LiNDa to guide an acoustic wave. An optical image of the ridge waveguide device is shown in Fig. S4(C). A pair of interdigitated transducers (IDTs) are patterned on the LN to convert microwaves to acoustics and vice versa. We design the waveguides with help from COMSOL's finite element analysis method. There are four fundamental SAW modes of the waveguides as shown in Fig. S4(A). The simulated eigenfrequency of each fundamental acoustic waveguide mode is shown in Fig. S4(B). The width of the waveguide in Fig. S4(C) and (D) is 800 nm, and the wavelength is 800 nm. The waveguide thickness is 200 nm, and the remaining layer thickness is 100 nm. The acoustic wave in the waveguide is propagating along the y-axis of lithium niobate. In our fabricated design, we define an IDT finger pair period (i.e. SAW wavelength) of 800 nm and an electrode length of 4 $\mu m$. Aa 30-$\mu m$ long tapering region guides the acoustic wave into the acoustic waveguide.

We start the fabrication by transferring a 300-nm LN membrane to a diamond substrate. The waveguide patterns are first defined with EBL using Zep520A (a positive electron beam lithography resist) and then transferred to LN by etching down 200 nm using argon ion etching. This is followed by a redeposition clean in a SC1 solution. The electron beam resist is stripped in a piranha solution (3:1 98% sulfuric acid and 40% hydrogen peroxide). A scanning electron microscope image of the curved acoustic waveguide is shown in Fig. S4(D). Finally, the metal IDTs are defined by EBL using PMMA 950 C4 as the resist. We evaporate 110 nm of aluminum and create the electrodes by a lift-off process in remover PG at 80 degrees Celsius. The wirebond pads are defined with another EBL write followed by evaporation of 10 nm titanium and 300 nm gold and another identical lift-off process.

The device is characterized with frequency-domain transmission and reflection measurements as shown in Fig. S4(E). The transmission measurement is gated to effectively suppress microwave crosstalk. The measured transmission exhibits a resonance at 6.2 GHz, with a transmission magnitude of -57 dB. The observed high transmission loss is attributed to the fabrication complexity and mode mismatch between the IDT region and the waveguide region. We will discuss each contributing factor in detail.

**Fabrication complexity**: A post-etch clean step using SC1 solution is crucial to remove redeposited residues following the argon ion etching of LN. However, this cleaning process can induce delamination of the LN membrane from the diamond substrate. To address this, a 10-nm chromium layer and a 350-nm gold layer are deposited along the membrane perimeter to mechanically anchor the LN to the diamond substrate. The metal layers are subsequently removed using gold and chromium etchants prior to the patterning of IDTs. While effective, the use of the metal clamp layers introduces additional fabrication complexity. Furthermore, the multiple chemical processes involved during device fabrication may contribute to increased surface roughness and potential degradation of acoustic performance.

**Mode mismatch**: The IDTs launch acoustic waves into the slab region, where a mismatch arises between the slab mode and the guided waveguide mode. Extending the taper region and incorporating the anisotropic properties of LN into the design could enable more adiabatic mode conversion, thereby improving coupling efficiency between the two modes. However, the current taper length is constrained by the limited size of the LN membrane (~ 200 $\mu$m).

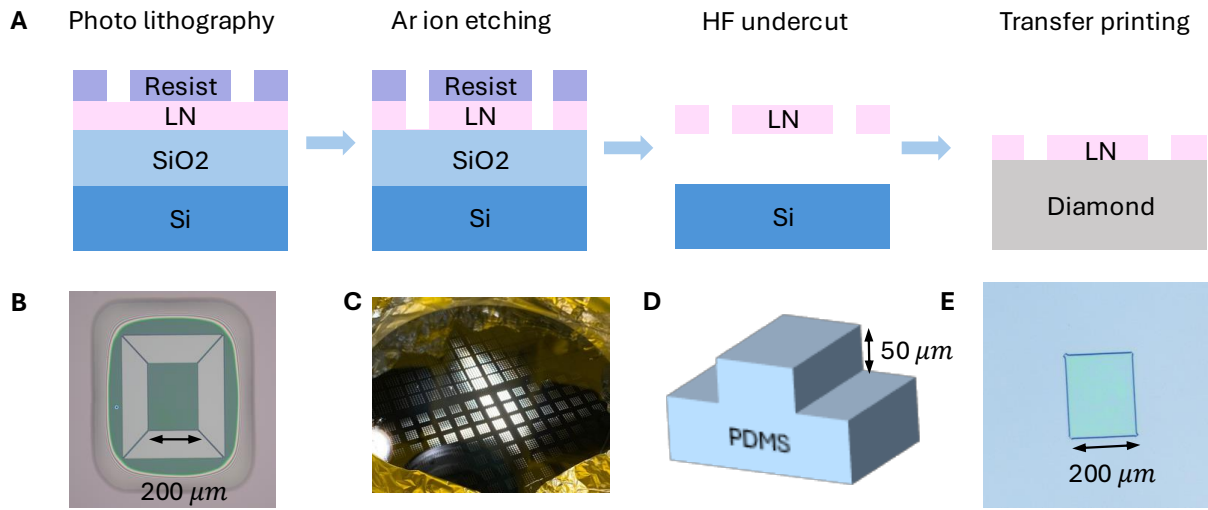

**Fig. S1. Fabrication process used to realize the LiNDa platform.** **(A)** Process flow for preparing and transferring thin-film lithium niobate membranes. **(B)** Optical image of a suspended LN membrane with thin tethers. The green contrast highlights the undercut region. **(C)** Patterned polydimethylsiloxane (PDMS) stamps on a silicon wafer. **(D)** The schematic of a single PDMS cell used in the transfer. The top part is 50-$\mu m$ thick and its lateral dimensions are 295 $\mu m$ by 463 $\mu m$, which are large enough to cover the whole suspended LN membrane area and break the tether when lifting up the PDMS. **(E)** Optical image of a transferred 600-nm thick LN membrane transferred onto a diamond substrate. The dimension of the LN membrane is 200 $\mu m$ by 250 $\mu m$.

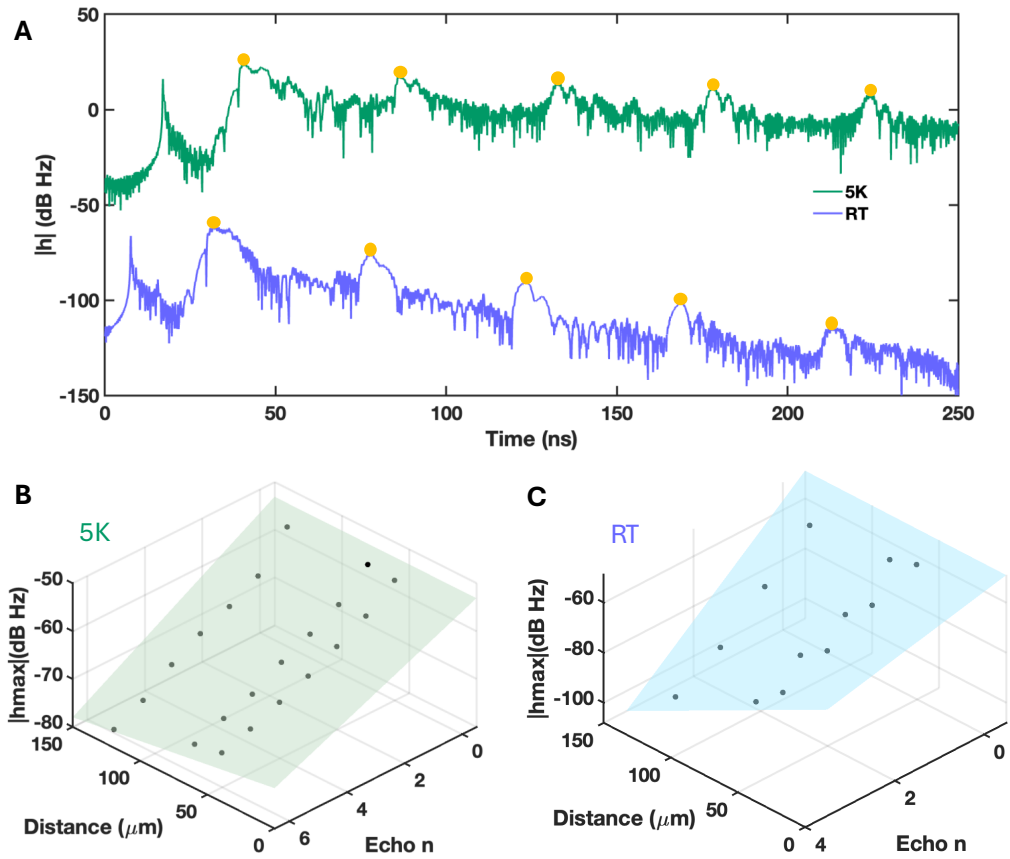

**Fig. S2. Analysis of acoustic propagation loss in planar SAW devices. (A)** Time domain response $|h|$ for a 130-$\mu m$ long device at 5K and room temperature (RT). The dots are the peaks of each echo and are measured to be $|h_{max}|$. **(B)** and **(C)** A linear fit is applied to the maximum impulse response $|h_{max}|$ to extract the propagation loss at 5K and at room temperature, respectively. The propagation loss is extracted to be 35.2 dB/mm at room temperature and 3.2 dB/mm at 5K.

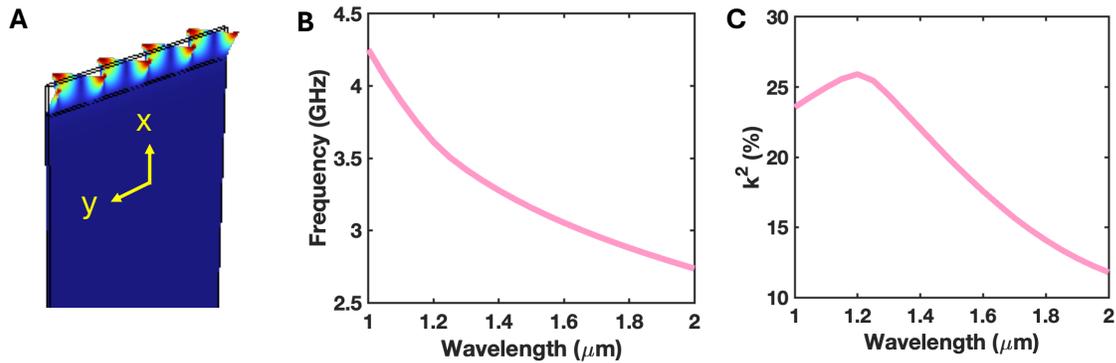

**Fig. S3. Simulations of SAW modes and spin-phonon coupling strength. (A)** Simulated shear mode in a 600-nm x cut LN on diamond, obtained using the finite-element analysis method in COMSOL. The acoustic wave propagates along the y axis of lithium niobate. **(B)** Simulated eigenfrequency of the shear mode when the electrode on the surface is electrically shorted as the wavelength varies. **(C)** Simulated electromechanical coupling coefficient $k^2$ as the wavelength varies. The highest $k^2$ is around 25%.

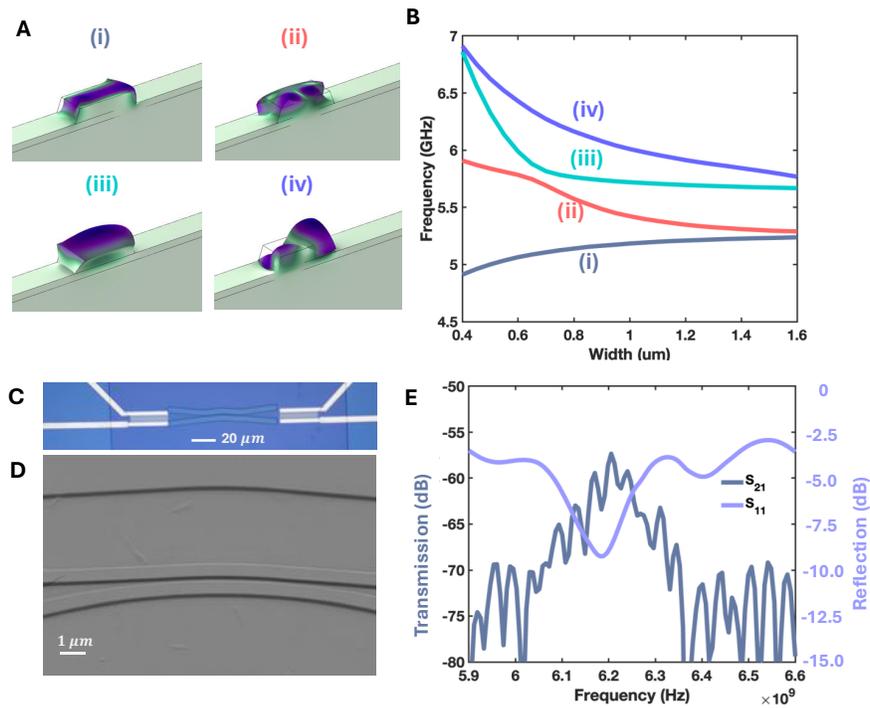

**Fig. S4. SAW ridge-waveguide devices on LiNDa platform.** (**A**) Simulated waveguide modes using 300-nm thick xcut LN on diamond, obtained using the finite-element analysis method in COMSOL. The acoustic wave in the waveguide is propagating along the y axis of the LN. (**B**) The simulated eigenfrequency of four fundamental modes as waveguide width increases. The wavelength is 800 nm. (**C**) Optical image of a curved acoustic waveguide. The pattern is defined by positive resist and transferred to LN by argon ion milling. (**D**) The scanning electron microscope (SEM) image of a curved acoustic waveguide. (**E**) Frequency-domain transmission $S_{21}$ and reflection $S_{11}$ measurement. The transmission spectrum is gated to effectively reduce microwave crosstalk.